\documentclass[%
 reprint,
nofootinbib,
 amsmath,amssymb,
 aps,
 pra,
 twocolumn
]{revtex4-2}
\usepackage{graphicx}
\usepackage{dcolumn}
\usepackage{bm}
\usepackage{braket}
\usepackage{slashed}
\usepackage{hyperref}
\usepackage{csquotes}
\usepackage{relsize}
\usepackage{mathrsfs}
\usepackage{tikz}
\usepackage{bm}
\usepackage{euscript}
\usepackage{soul}

\newcommand{\be}{\begin{equation}}
\newcommand{\ee}{\end{equation}}
\newcommand\beq{\begin{eqnarray}}
\newcommand\eeq{\end{eqnarray}} 
\newcommand\eqn[1]{\label{eq:#1}} 
\newcommand\eq[1]{eq.~(\ref{eq:#1})} 
\newcommand\Eq[1]{Eq.~(\ref{eq:#1})} 
\newcommand\fig[1]{Fig.~\ref{fig:#1}} 

\newcommand{\bfe}{{\mathbf e}}

\newcommand{\bfns}{{{\mathbf n}^\star}}

\newcommand{\bfn}{{\mathbf n}}

\newcommand{\olatt}{{L}}
\newcommand{\bfell}{{\boldsymbol  \ell}}
\newcommand{\bfells}{{{\boldsymbol  \ell}^\star}}
\newcommand{\dlatt}{{L^\star}}

\newcommand{\bmc}{{\bm c}}
\newcommand{\bmcs}{{\bm c}^\star}
\newcommand{\bmp}{{\bm p}}
\newcommand{\bmps}{{{\bm p}^\star}}

\newcommand{\bfnu}{{\boldsymbol  \nu}}

\newcommand{\MCH}{\EuScript{H}}
\newcommand{\MCHs}{{\EuScript{H}^\star}}
\newcommand{\EA}{{\mathscr {A}}}
\newcommand{\EH}{\mathscr{H}}
\newcommand{\EP}{{\mathscr {P}}}
\newcommand{\EQ}{{\mathscr {Q}}}
\newcommand{\EU}{{\mathscr {U}}}
\newcommand{\EW}{{\mathscr {W}}}

\newcommand{\CE}{{\cal E}}

\newcommand{\Zint}{{\mathbb{Z}}}

\usepackage[section]{placeins}

\begin{document}

\preprint{INT-PUB-18-034}

\title{Gauss's Law, Duality, and the Hamiltonian Formulation of U(1) Lattice Gauge Theory}
\author{David B.~Kaplan}
\email{dbkaplan@uw.edu}
\affiliation{%
Institute for Nuclear Theory, Box 351550, University of Washington, Seattle, Washington 98195-1550, USA
}%
\author{Jesse R.~Stryker}
\email{stryker@uw.edu}
\affiliation{%
Institute for Nuclear Theory, Box 351550, University of Washington, Seattle, Washington 98195-1550, USA
}%


\begin{abstract}
  Quantum computers have the potential to explore the vast Hilbert space of entangled states that  play an important role in the behavior of strongly interacting matter.
  This opportunity has motivated reconsidering the Hamiltonian formulation of gauge theories, with a suitable truncation scheme to render the Hilbert space finite-dimensional.
  Conventional formulations lead to a Hilbert space largely spanned by unphysical states;
  given the current inability to perform large scale quantum computations, we examine here how one might restrict wave function evolution entirely or mostly to the physical subspace.
  We consider such constructions for the simplest of these theories containing dynamical gauge bosons---U(1) lattice gauge theory without matter in $d=2,3$ spatial dimensions---and find that electric-magnetic duality naturally plays an important role.
  We conclude that this approach is likely to significantly reduce computational overhead  in $d=2$ by a reduction of variables and by allowing one to regulate magnetic fluctuations instead of electric.   The former advantage does not exist in $d=3$, but the latter might be important for asymptotically-free gauge theories.

  \end{abstract}
\maketitle 
Wilson's path integral construction provides a nonperturbative definition of lattice gauge theory and an efficient computational tool for some types of calculations.
However, even for quantum chromodynamics (QCD) many properties elude a first-principles understanding, such as the phase diagram at finite chemical potential, real-time dynamics, topological properties, and the structure of all but the lightest nuclei.
Such investigations entail  sign problems that are exponentially hard to solve on a classical computer.
Quantum computers offer hope for surmounting these obstacles and a number of papers have proposed using the Kogut-Susskind \cite{kogut.susskindHamiltonianFormulation75} lattice Hamiltonian $H_\text{KS}$ as a starting point for the study of gauge theories, introducing a cutoff on the electric field (in addition to the finite lattice spacing) in order to render the Hilbert space $\MCH$  finite-dimensional \cite{byrnes.yamamotoSimulatingLattice06,zohar.burrelloFormulationLattice15,wieseQuantumSimulating14} (for discussions of Hamiltonian lattice gauge theory, see \cite{creutzQuarksGluons84,smitIntroductionQuantum02}).
The vast majority of states in $\MCH$ are unphysical;
the physical space is limited to those obeying Gauss's law, which we will call $\MCH_\text{phys}\subset\MCH$.

There are a couple of drawbacks to this approach which we address here:
(i) It appears preferable to work entirely in $\MCH_\text{phys}$ if possible, in order to require fewer qubits and to avoid computational errors causing states initially in $\MCH_\text{phys}$ to evolve into the much larger space of unphysical states;
(ii) A cutoff on electric fields is appropriate for strong coupling, for which electric fluctuations are suppressed, but is not ideal for weak coupling, such as one would encounter in the continuum limit for any gauge theory in $d<3$, or asymptotically-free theories in $d=3$, where $d$ is the spatial dimension. Instead, a cutoff on magnetic fluctuations would likely be a more efficient regulator, allowing one to approach the continuum limit with a smaller Hilbert space.
In this paper we examine these issues in two of the simplest gauge theories -- 
U(1) theories without matter in $d=2$ and $d=3$ -- and find that both concerns lead directly to a formulation of the electromagnetic dual theory.
Dualities in lattice gauge theories are as old as lattice gauge theory itself
\cite{%
balian.drouffe.eaGaugeFields75,%
banks.myerson.eaPhaseTransitions77,%
fradkin.susskindOrderDisorder78,%
thooftPhaseTransition78,%
yoneyaTopologicalExcitations78,%
korthalsaltesDualityGauge78,%
greenDiscreteAbelian78,%
elitzur.pearson.eaPhaseStructure79,%
horn.weinstein.eaHamiltonianApproach79,%
ukawa.windey.eaDualVariables80,%
orlandDualityNonabelian80,%
kogutIntroductionLattice79,%
savitDualityField80%
}
and experienced a surge of interest during the early days of Hamiltonian lattice gauge theory.
While the dual theories encountered below are not of direct physical interest, they are simple enough to clearly illustrate a subset of the issues that must be faced when simulating U(1) gauge theories with matter, or non-Abelian gauge theories.


\section{U(1) Hamiltonian and Hilbert space} 


The  Hamiltonian for a U(1) gauge theory in the continuum is $ \hat H =(1/2) \int d^dx\,  (\hat E^2 + \hat B^2)$, where
 the electric field operator  $\hat E_i$ is the conjugate momentum for the vector potential $\hat A_i$. Here we consider compact U(1) gauge theory formulated on a spatial lattice $\olatt$ with lattice spacing $a_s$, periodic boundary conditions, and coordinates $\{\bfn,\bfell,\bmp,\bmc\}$ for sites, links, plaquettes, and cubes, respectively.   Compact U(1) theory is interacting at finite $a_s$; because time and space are treated asymmetrically, there are two coupling constants $g_{t,s}$ which must be independently renormalized, with dimensionless couplings  defined as
$
\tilde g^2_{t,s} = a_s^{3-d}g^2_{t,s}
$. The continuum limit is equivalent to $\tilde g^2_{t,s}\to 0$ for $d<3$ (as well as   for asymptotically-free non-Abelian gauge theories in $d=3$).  
We fix $A_0=0$ gauge,  and replace the vector potential $ {\bf A}(x)$ by a unitary operator $\hat U_{\bfell}= \exp( -i g_s a_s \hat A_{\bfell})$  on every link;
$\hat U_{\bfell}$ can be thought of as the coordinate operator for  a particle moving on the group manifold.
The space $\MCH$ can be represented in the coordinate basis of product states $\otimes_\bfell\ket{U_\bfell}$, where $\ket{U_\bfell}$  at each link $\bfell$ is an eigenstate of   $\hat U_\bfell$ with eigenvalue $U_\bfell$, which is a phase.
Alternatively, one can work in the momentum basis, which diagonalizes the electric field $E_{\bfell} $ also residing on the links.
The rescaled electric field
  \beq
\hat\CE_{\bfell} \equiv \frac{a_s^{\frac{d+1}{2}}}{\tilde g_s}\,\hat E_{\bfell} 
\eeq
satisfies the  commutation relation
\beq
\left[\hat \CE_{\bfell}, \hat U_{\bfell'}\right] =   \delta_{\bfell,\bfell'} \hat U_{\bfell}\ ,
\eqn{ecom}\eeq
and has integer eigenvalues $\varepsilon_\bfell$, analogous to the angular momentum of a particle on a circle.  
  $\MCH$ can then be represented in the electric field basis of product  states $\otimes_\bfell\ket{\varepsilon_\bfell}$ and regulated in a gauge-invariant way  by restricting fluctuations of the electric field,  $|\varepsilon_\bfell|\le N$ for some cutoff $N$ \cite{byrnes.yamamotoSimulatingLattice06}.

Our starting point for the lattice Hamiltonian is $ \hat H=\hat H_E + \hat H_B$, with
\beq
\hat  H_B &=&\frac{1}{2a_s}\left[\frac{1}{  \tilde g_s^2} \sum_{\bmp}\,\left(2-\hat P_{\bmp} - \hat P^\dagger_{\bmp}\right) \right], \cr
 \hat H_E &=& \frac{1}{ 2 a_s}\left[ \frac{ \tilde g_t^2}{\xi^2} \sum_{\bfell}\, \left(2-\hat Q_{\bfell} - \hat Q^\dagger_{\bfell}\right) \right],
\eqn{H} \eeq
where we define
\beq
 \hat Q_{\bfell}  \equiv e^{i \xi\hat \CE_{\bfell}}\ ,\qquad \hat P_{\bfn,ij}  \equiv \hat U_{\bfn,i} \hat U_{\bfn+ \bfe_i,j}\hat U^\dagger_{\bfn+\bfe_j,i} \hat U^\dagger_{\bfn,j} .
\eqn{QPdef}\eeq
Here $\hat  H_B$ is conventional with   $\hat P_{\bmp}$ being the usual plaquette operator, but  in  $\hat  H_E$ we have introduced the dimensionless parameter $ \xi$ for convenience, where
\eq{H} yields the Kogut-Susskind Hamiltonian $\hat{H}_\text{KS}$  in the limit $\xi\to 0$.
This is similar to the Hamiltonian for Z($N$) gauge theory in \cite{horn.weinstein.eaHamiltonianApproach79}.
The parameter $a_t\equiv \xi a_s$ can be thought of as a ``temporal lattice spacing,'' and  additional irrelevant terms subleading in $a_t$ could be added, but the above symmetric form suits our purposes best.  \Eq{ecom} implies that $\hat U$ acts as a raising operator for the electric quantum number, and can be expressed in the electric field basis as $\hat U= \sum_\varepsilon \ket{\varepsilon+1}\bra{\varepsilon} $.
The action of $\hat P$, therefore, is to create an oriented loop of unit electric flux around the edge of the plaquette, while $\hat P^\dagger$ creates a unit loop in the opposite direction.
At the same time, $\hat P$ measures magnetic field, the phase of its eigenvalue being the magnetic flux through the plaquette to leading order in  $a_s$.
The above form for $\hat H$ is bounded and written as a sum of unitary operators, which may be convenient for simulation by quantum walks \cite{low.chuangHamiltonianSimulation19}.  

Note that fluctuations in the magnetic field are large at strong coupling, while electric fluctuations are large at weak coupling.
This is similar to the case of a harmonic oscillator with mass $m$ and spring constant $k$, where    $\langle \hat x^2\rangle\propto 1/\sqrt{k m}$,  while $\langle \hat p^2\rangle\propto  \sqrt{k m}$, the operators  $\hat x$, $\hat p$ being analogues of $\hat B, \hat E$ respectively, while $m\sim 1/\tilde g_t^2$ and $k\sim 1/\tilde g_s^2$.
In other words, this analogy suggests $\langle E^2 \rangle \sim (\tilde{g}_t \tilde{g}_s)^{-1}$.
Thus, regulating the theory with a cutoff on electric field values may be a costly choice for  gauge theories in $d<3$, and asymptotically free gauge theories in $d=3$, as approaching the continuum limit would require an ever-increasing truncation level. 

The physical subspace $\MCH_{\text{phys}}\subset\MCH$  consists of those states obeying the Gauss law constraint  $\vec\nabla\cdot \vec E=0$, i.e., those states invariant under  spatial gauge transformations.
On the lattice, the analogue constraint is that at each lattice site the product of the $\hat Q$ on each outgoing link and $\hat Q^\dagger$ on each incoming link must equal the unit operator:
\beq
\left(\prod_{\bfell\text{ into }\bfn} \hat Q_{\bfell } \, \prod_{\bfell\text{ out of }\bfn} \hat Q^\dagger_\bfell  - \hat{\mathbf 1} \right) \ket{\text{phys}} = 0 \ .
\eqn{Gauss}
\eeq
Most states in $\MCH$ violate \eq{Gauss} and  are unphysical, and therefore simulating Hamiltonian evolution in $\MCH$ will use more qubits on a quantum computer than physically necessary.
To better understand this constraint, consider the lattice  $\olatt$  with periodic boundary conditions in $d=2,3$ dimensions with $n$ sites, and therefore $\ell = nd$ links, $p=nd(d-1)/2$ plaquettes, and $c=nd(d-1)(d-2)/6$ cubes.
The Hilbert space $\MCH$ is characterized by the eigenvalues of the $\ell$ electric field variables, $ \hat Q_{\bfell } $.
States fall into topological sectors labeled by an integer-valued $d$-tuple, $\bfnu=(\nu_1,\ldots, \nu_d )$ designating $\nu_i$ units of electric flux wrapping around the $\bfe_i$ direction of the lattice.
For a given topological sector we have $(n+d-1)$ constraints on the  $\ell=nd$ electric field variables: $(n-1)$ constraints from Gauss's law  and $d$ from fixing the topology.
Therefore there are $[nd-(n+d-1)]=(n-1)(d-1)$  variables to describe physical states in a particular topological sector.
If we place a cutoff on electric field values to regulate the theory, and assume $n\gg 1$, then the minimum number  of qubits required to describe  $\MCH_{\text{phys}}$ will scale as $(d-1)/d$ times the minimum number required for  $\MCH$;
this ratio is expected to be significantly smaller for non-Abelian theories.

The benefit of restricting a computation to $\MCH_{\text{phys}}$ is not only in reduction of qubits, but also in ensuring that computational errors do not propagate states into the unphysical part of $\MCH$, a process that would look like violation of charge conservation. 
\begin{figure}[t]
\includegraphics[width = 0.2\textwidth]{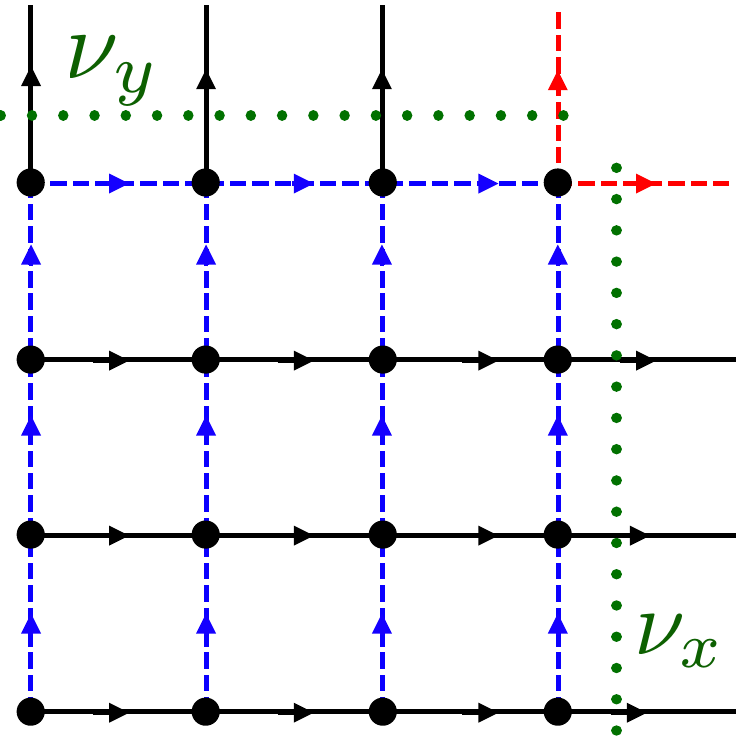}
\caption{\it An $n=16$ site lattice with periodic boundary conditions in $d=2$ with $nd=32$ links.
  The $(n-1)=15$ links on a maximal tree (dashed blue) are eliminated via Gauss's law.
  The $d=2$  links in dashed red are eliminated by constraining the net electric flux through the dotted green lines to equal the topological quantum numbers $\nu_{x,y}$.
  The remaining $(n-1)(d-1)=15$  black links represent the  physical variables of the theory.
  This procedure generalizes to arbitrary $d, n$.}
\label{fig:st}
\end{figure} 
A brute-force approach  for restricting  $\MCH\to \MCH_{\text{phys}}$ is to eliminate the constrained variables in the quantum theory by the   procedure illustrated in  \fig{st}:
(i) Define a maximal tree on the lattice, with $(n-1)$ links;
(ii) eliminate the $\ket{\varepsilon_\bfell}$ states from $\MCH$ for each link in the tree;
(iii) set  $\hat U_\bfell \equiv 1$  in $\hat H$ for each eliminated link;
(iv) recursively solve for  the $\hat Q_\bfell $ in $\hat H$  at each eliminated link, in terms of the  $\hat Q$'s on the free links;
(v) remove the final $d$ links by enforcing a fixed topology $\bfnu$.
The fourth step involves   defining  $\hat Q$ operators on each of the tree links as the appropriate product of the other  $\hat Q,\hat Q^\dagger$ operators  meeting at the same   vertex, beginning at the ends of the tree branches;
fixing the topology (step five) can be easily done at the border of the maximal tree, as indicated in \fig{st}. 
The resulting $\MCH_{\text{phys}}$ on a given sector $\bfnu$ is written using products of the $\ket{\varepsilon_{\bfell}}$ states over each of the $(n-1)(d-1)$ free links;
a heavy price is paid, however, in the loss of locality and discrete translational invariance of the resulting Hamiltonian. We next describe an alternative procedure, which leads directly to a duality transformation.


\section{U(1) dual formulation}


States in the gauge-invariant Hilbert space can be constructed by repeatedly applying the gauge invariant plaquette operators to the strong coupling vacuum $\ket{0}$ \cite{kogut.susskindHamiltonianFormulation75}, as well as Wilson loop operators with  nontrivial winding number around the lattice, where $\ket{0}$ is the state annihilated by the electric field.
First we define the Polyakov loop operators $\hat W(C_i)$ to be the product of oriented $\hat U$ link operators along a closed loop $C_i$ that wraps around the lattice in the $\bfe_i$ compact direction.
We then define the state
\beq
\ket{\bfnu} \equiv \prod_{i=1}^d \left(\hat W(C_i)\right)^{\nu_i} \ket{0} \ ,\quad {\nu_i}\in\Zint\ .
\eqn{Wloops}\eeq
All the physical states within a topological class are then created by acting on $\ket{\bfnu}$ with powers of plaquette operators: 
\beq
\ket{\EA}_\bfnu  =\prod_{\bmp}\left( \hat P_{\bmp}\right)^{\EA_{\bmp}}\ket{\bfnu} \ ,\  \EA_{\bmp}\in\Zint \ ,
\eqn{Adef}\eeq
where $\bmp$ runs over all $p$ plaquette coordinates. It is evident that $\ket{\EA}_\bfnu \in \MCH_{\text{phys}}$ for all $\EA$ and $\bfnu$ since both the $\hat W$ and $\hat P$ operators are gauge-invariant, each producing only closed loops of electric flux. It is also easy to see that any of the $\otimes_\bfell\ket{\varepsilon_\bfell}$ basis states obeying Gauss's law can be written in this form.
The particular choice of the $C_i$ paths is unimportant, since two such paths can be deformed into each other by the application of plaquette operators.

The fact that the dynamics of this theory is expressible in terms of an integer-valued field could also be expected from early studies of field theoretic partition functions with U(1) symmetries \cite{banks.myerson.eaPhaseTransitions77,savitDualityField80}.
Above, the dual variables arise as a consequence of directly analyzing the form of states that are connected to the strong-coupling vacuum by powers of $\hat{H}_B$.

A problem remains, and that is that the $\ket{\EA}_\bfnu$ states are an overcomplete basis for $\MCH_{\text{phys}}$, seeing as a state in a particular topological sector depends on $  p =  nd(d-1)/2$ variables instead of the required $(n-1)(d-1)$.
The number of redundant $\EA$ variables is therefore $R=(d-1)[1 +(d-2)n/2]$.
For $d=2$, the redundancy is $R=1$, independent of the number of sites $n$; for $d=3$, $R=2+n$, scaling with the volume of the lattice.
These redundancies arise because the product of plaquette operators around any closed surface is an identity operation,  expressing a discretized form of $\oint  d^2 x \, \vec{B} \cdot \hat{n} = 0 $ on the toroidal surface;
$R$ simply counts the number of independent closed surfaces.
We will deal with the redundancy by treating all of the $\ket{\EA}_\bfnu$ states as independent, then subsequently  imposing this magnetic Gauss law constraint.

The  action of the Hamiltonian \eq{H} on the $\ket{\EA}_\bfnu$ states is simple to characterize:
$\hat H_B$ applies plaquette operators to the state, and therefore either raises or lowers $\EA_{\bmp}$ by one.
$\hat H_E$ measures the electric field, which at each link is determined by differences between the $\EA_{\bmp}$ for the plaquettes the link borders --- with a possible additional contribution from the Polyakov loop in \eq{Wloops} if the link lies along one of the $C_i$ curves.
$\hat H_E$ therefore looks like a finite difference operator acting on  $\EA_{\bmp}$.  The behavior of $\hat{H}_B$ and $\hat{H}_E$ can be most naturally described in terms of operators on the dual lattice.
We first discuss the simpler case of $d=2$,  where the  duality transformation maps the $\{\bfn,\bfell,\bmp\}$ coordinates of  $\olatt$   to $\{\bmp^\star,\bfell^\star,\bfns\}$, respectively, on the dual lattice $\dlatt$.
$\bfn$ sits at the center of plaquette $\bmp^\star$ and $\bfns$ sits at the center of $\bmp$, while $\bfell^\star$ and $\bfell$ intersect each other;
we adopt a convention where the $x$-links of the two lattices are oriented anti-parallel to each other, while the $y$-links are parallel.
 By ignoring the redundancy in our  definition of  $\ket{\EA}_\bfnu$ in \eq{Adef}, we can treat $ \EA_\bfns$ as an independent integer-valued variable on each site and use product states $\otimes_\bfns\ket{\EA_\bfns}$ as a basis for a Hilbert space $\MCHs$.
 In terms of these states, we can define  the two local coordinate and shift operators, $\hat \EU_\bfns$ and $\hat \EQ_\bfns$, living on sites of the dual lattice as
 \beq
\hat \EU_\bfns  &=& \sum_{\EA_\bfns} \ket{\EA_\bfns} e^{i\xi\EA_\bfns}\bra{\EA_\bfns}\ ,\cr
\hat \EQ_{\bfns}  &=&\sum_{\EA_\bfns}  \ket{\EA_\bfns+1} \bra{\EA_\bfns}\ .
\eqn{dualQPdef}\eeq
For a given topological sector $\bfnu$, the matrix elements of the Hamiltonian $\hat H$ of \eq{H}  between the $\ket{\EA}_\bfnu$ states  are reproduced then by the dual Hamiltonian $\hat \EH_{\bfnu} $ on $\dlatt$,
\begin{equation}
\begin{aligned}
\hat \EH_{\bfnu}  =  \frac{1}{2 a_s} \sum_{\bfns}\biggl[&\frac{1}{\tilde g_s^2} \left(2-\hat \EQ_{\bfns} -\hat \EQ^\dagger_{\bfns}\right) \cr &-    \frac{\tilde g_t^2 }{\xi^2}  a_s^2\hat{\EU}_{\bfns}^\dagger \Delta  \hat{\EU}_{\bfns}\
 \biggr] \qquad (d=2).
\end{aligned}
\eqn{Hd2}
\end{equation}
  In this expression, $\Delta $ is a discrete covariant Laplacian  $\Delta = D_i^+ D_i^- $, where 
$D_i^+$ are the difference operators
\beq
D^+_1 F_{\bfns} &=& ( \EW_{\{\bfns, \bfns-\bfe_1\}} F_{\bfns-\bfe_1}-F_{\bfns}   )/a_s\ , \cr
D^+_2 F_{\bfns} &=& ( \EW_{\{\bfns,\bfns+\bfe_2\}} F_{\bfns+\bfe_2} -  F_{\bfns} )/a_s\ , 
\eeq
$D^-_i \equiv  -\left(D^+_i \right)^\dagger $, and the discrete vector gauge field $\EW$ accounts for the topological charges $\bfnu$:
\beq
\EW_{\bfell^\star} =
\begin{cases}e^{i \xi \nu_i }, & \text{if }\bfell \in C_i \, ;\\
  1, & \text{otherwise} \, ; \\
\end{cases}
\eeq
$\bfell^\star$ being the link dual to $\bfell$. 
Note that $D_1^+$ is a derivative in the $-\bfe_1$ direction because on $\dlatt$ we have oriented the $x$-links anti-parallel to those on $\olatt$, unlike the $y$-links, which are parallel.
The gauge symmetry associated with $\EW$ reflects the equivalence of constructions based on different $C_i$ paths for the Polyakov loops in \eq{Wloops}.

The first term in \eq{Hd2} arises from $\hat H_B$, while the second arises from $\hat H_E$, and we see that the roles of the two have been reversed:  $\hat H_B$ becomes an operator that translates the value of the  dual field $\EA$, while $\hat H_E$ measures spatial derivatives of $\EA$.
The discrete gauge field $\EW$ corresponding   to the topological electric fields of the original theory  seems to have no analogue in the original theory, but that is simply because we did not build in topological magnetic field loops; to do so would require a field analogous to $\EW$ added to the original Hamiltonian $\hat H$.   
 
As mentioned above, in $d=2$ there is one redundant variable arising from the fact that $\prod_\bmp \hat P_\bmp = \hat{\mathbf 1}$.   Thus the restriction to $\MCH_\text{phys}\subset\MCHs$ requires applying the single constraint on physical states
\beq
\left(\hat\EQ_{\dlatt} -\hat{\mathbf 1}\right)\ket{\EA}_{\bfnu} =0\ ,\qquad \hat\EQ_{\dlatt}\equiv \prod_\bfns  \hat\EQ_\bfns\ .
\eqn{2dcon}
\eeq
This constraint  
can be solved by setting $\EA=0$ at a single site $\bfns$ and equating $ \hat \EQ$ at that site to the product of  $\hat \EQ^\dagger$ over all the other sites --- again at the cost  of sacrificing locality and discrete translational invariance.  A more attractive alternative is to work directly in $\MCHs$ and simply use an initial wave function that satisfies \eq{2dcon}.  Unlike in the conventional formulation, where 
the number of constraints scales with the number of lattice sites, here with only a single unphysical variable, the problems of  constructing the initial state obeying the constraint ---  or of subsequently becoming ``lost in space" due to computational error --- should be vastly diminished compared to simulations in the original space $\MCH$ subject to \eq{Gauss}.  Because there is one $\hat\EQ_\bfns, \hat \EU_\bfns$ variable pair per site on $\dlatt$, as compared with two $\hat Q_\bfell, \hat U_\bfell$ variable pairs   per site on $\olatt$, we see the expected $(d-1)/d=1/2$ reduction in degrees of freedom, which should correspond to a similar reduction in the number of qubits required to characterize the system. However, for this statement to be concrete, we first have to discuss regulating $\MCHs$ to make it finite-dimensional. 

To regulate the dual theory in $d=2$ one cannot simply limit $\EA_{\bfns}$ to lie in the finite range $-N \le \hat \EA_{\bfns}\le N$, taking    $N\to\infty$ in the continuum limit:
 the operator $   \hat\EQ_{\dlatt}$ shifts the $\hat\EA_\bfns$ field uniformly so that the constraint  \eq{2dcon} cannot hold in a space spanned by  eigenstates of the $\hat\EA_\bfns$ with finite eigenvalues.
Instead, one can regulate the eigenvalues of the unitary $\EQ_\bfns$ operators, equivalent to placing a cutoff on magnetic field fluctuations in the original theory.
The regulated Hamiltonian will then commute with the constraint, and an initial wave function chosen to satisfy the constraint \eq{2dcon} will continue to do so as it evolves.
Therefore, in $d=2$, there are several advantages to simulating $\hat \EH_\bfnu$ on a quantum computer instead of $\hat H$: (i) the variables are scalars, rather than vectors, reducing the number of degrees of freedom by half; (ii) there is a single redundant variable, rather than the $(n-1)$ unphysical variables in the conventional formulation; (iii) it is natural to regulate magnetic fluctuations rather than electric, which is likely to converge more efficiently to the continuum limit.

We now turn to the problem of constructing the $d=3$ Hamiltonian for the $\ket{\EA}_\bfnu$ states of \eq{Adef}.  As for $d=2$, this naturally leads to a duality transformation, interchanging the coordinates for sites, links, plaquettes and cubes   from $\olatt$ to $\dlatt$ as $\{\bfn,\bfell,\bmp,\bmc\}\leftrightarrow \{\bmcs,\bmps,\bfells,\bfns\}$.
In particular, the   plaquettes $\bmp$ on $\olatt$ get mapped to the links $\bfells$ on $\dlatt$ piercing them in the direction opposite to their normal vectors, so that $\dlatt$ is parity inverted relative to $\olatt$.  Therefore the plaquette variable $\EA_\bmp$ on $\olatt$ gets mapped to a dual vector field $\EA_\bfells$ living on the links of $\dlatt$, unlike in $d=2$ where a scalar $\EA_\bfns$ lives on sites.  We can then define link operators $\EU_\bfells$ and $\EQ_\bfells$ operators exactly as in \eq{dualQPdef}, and the dual Hamiltonian is computed to be
\begin{equation}
\begin{aligned}
  \hat \EH_{\bfnu}  =  \frac{1}{2 a_s}&\biggl[\sum_\bfns \frac{1}{\tilde g_s^2} \left(2-\hat \EQ_\bfells -\hat \EQ^\dagger_\bfells\right) \cr &+   \frac{ \tilde g_t^2}{\xi^2}   \sum_\bmps \left( 2- \left(\EW_\bmps \hat\EP_\bmps+ \text{h.c.}\right) \right)
  \biggr] \quad (d=3) .
\end{aligned}
\eqn{Hd3}
\end{equation}
In this expression,
 $\EP_\bmps$ is the plaquette operator on $\dlatt$ constructed out of $\EU_\bfells$'s in the same way $ P_\bmp$ is constructed from $ U_\bfell$'s in \eq{QPdef},
while $\EW_\bmps$ is a phase that is nontrivial whenever the topological  electric field loops $C_i$  on $\olatt$ pierce the $\bmps$ plaquette on $\dlatt$,
\beq
\EW_{\bmps} =
\begin{cases}e^{i\xi \nu_i }, & \text{if }\bfell \in C_i \, ;\\
  1, & \text{otherwise} \, ;\\
\end{cases}
\eeq
$\bfell $ being the link dual to $\bmps$.

In   $d=3$ the $\ket{\EA}_\bfnu$ states  in  \eq{Adef} are again overcomplete, but  the problem is more severe than in $d=2$ as the product of plaquette operators on the surface of any cube $\bmc$ in $\olatt$ should be an identity transformation, the number of cubes scaling with $n$.   The constraint on the dual lattice to remove this degeneracy is (unsurprisingly) the dual of the electric Gauss law constraint \eq{Gauss}: the same equation with the substitution $\hat Q_\bfell\to   \hat \EQ_\bfells$. 
Thus, we see a ``conservation of difficulty'' between the original and dual theories for $d=3$, each form of the theory having a Gauss law constraint of identical form.  The    one advantage of the dual formulation common with the $d=2$ example is that  regulating the eigenvalues of the $\hat\EQ$ operators controls magnetic fluctuations, which we expect to be more efficient at weak coupling than a cutoff on the electric field.


\section{Conclusions}


We have focused here entirely on U(1) gauge theories without matter and have shown the consequences of defining these theories on the space of gauge-invariant states.  In particular, we found that this leads to a dual formulation subject to a magnetic Gauss law constraint.  This result can lead to a substantial reduction of variables in $d=2$, but not in $d=3$; in both cases though it offers the opportunity to regulate the theory by limiting magnetic fluctuations rather than electric, which is expected to be advantageous in approaching the continuum limit in $d=2$, or studying the weak field limit in $d=3$.  One can hope for a similar approach to regulating asymptotically-free gauge theories in $d=3$, for which the continuum limit is also at weak coupling.
Extending the analysis to include charged matter fields and non-Abelian gauge symmetries is complicated by the fact that not all gauge-invariant states in the theory can be written in the form \eq{Adef};
new progress has been made on the former since the initial preparation of this manuscript \cite{bender.zoharGaugeRedundancyfree20}, while much  previous  work on related issues for non-Abelian gauge theories exists  \cite{anishetty.sharatchandraDualityTransformation90,mathurLoopStates06,mathurLoopApproach07,raychowdhuryLowEnergy19} and could serve as a basis for quantum computations.
Understanding such theories better, and developing the tools to efficiently represent these theories on a quantum computer and extrapolate to the continuum theory, remain as fascinating theoretical challenges to be tackled before one can contemplate solving outstanding computational problems in QCD.

\begin{center}
  * * *
\end{center}

Subsequent to the submission of this paper, the authors of Ref.~\cite{haase.dellantonio.eaResourceEfficient20} presented results on the impact of different truncation schemes in Abelian lattice gauge theories as a function of the bare coupling.
That work analyzes (2+1)-dimensional U(1) theories in great detail, including fermionic matter, in both electric and magnetic representations.
They find that recovering the weak-coupling ground state wave function of a 2$\times$2 periodic lattice with high overlap requires a truncation level scaling like $\tilde{g}_s^{8/5}$ in a magnetic representation, as compared to $\tilde{g}_s^{-2}$ in an electric representation.
The former is clearly much more efficient on qubit resources at weak coupling, quantitatively confirming our qualitative arguments.
Their magnetic basis results are obtained by replacing the U(1) gauge group by its Z($N$) subgroups, however, a method that cannot be directly applied to non-Abelian gauge groups.
{This work was followed up by an experimental proposal for simulating 2D QED in near-term trapped ion quantum computers} \cite{paulson.dellantonio.eaSimulating2D20}{.

In addition, Refs.} \cite{bender.zoharGaugeRedundancyfree20} {and} \cite{bender.emonts.eaRealtimeDynamics20} {appeared, addressing dual variables for compact U(1) gauge theories that incorporate charged matter and that preserve translational invariance.}

\section{Acknowledgments}


We would like to thank N. Klco, J. Lombard, M. Savage,  L. Yaffe and for insightful conversations during the preparation of this manuscript.   This work was supported in part by DOE Grant No. DE-FG02-00ER41132 and by the Thomas L. and Margo G. Wyckoff Endowed Faculty Fellowship;  J.R.S. was also supported in part by the National Science Foundation Graduate Research Fellowship under Grant No.~DGE-1256082 and by the Seattle Chapter of the Achievement Rewards for College Scientists Foundation.


\bibliography {U1duality.bib}

\end{document}